\documentclass[aps,prl,floats,floatfix,superscriptaddress,twocolumn]{revtex4}%twocolumn
\usepackage{latexsym,amsmath}
\usepackage{amsbsy}
\usepackage[pdftex]{graphicx}
\usepackage{epstopdf}
\usepackage{amssymb}
\usepackage{epstopdf}
\usepackage[usenames]{color}
\usepackage{hyperref}
\usepackage{verbatim}
\usepackage{multirow,tabularx}
\usepackage[normalem]{ulem}
\usepackage[percent]{overpic}
\usepackage{booktabs}

\definecolor{olive}{RGB}{20,164,0}
\definecolor{orange}{RGB}{255, 155, 0}

\newcommand{\av}[1]{\langle #1 \rangle}

\newcommand{\FigPath}{./figs/}

\begin{document}

\title{Geometric randomization of real networks with prescribed degree sequence }

\author{Michele Starnini}
\affiliation{Data Science Laboratory, ISI Foundation, Via Chisola 5, 10126, Torino, Italy}
\author{Elisenda Ortiz}
\affiliation{Departament de F\'{\i}sica de la Mat\`eria Condensada, Universitat de Barcelona, Mart\'{\i} i Franqu\`es 1, 08028 Barcelona, Spain}
\affiliation{Universitat de Barcelona Institute of Complex Systems (UBICS), Universitat de Barcelona, 08028, Barcelona, Spain}
\author{M. \'Angeles Serrano}%
\affiliation{Departament de F\'{\i}sica de la Mat\`eria Condensada, Universitat de Barcelona, Mart\'{\i} i Franqu\`es 1, 08028 Barcelona, Spain}
\affiliation{Universitat de Barcelona Institute of Complex Systems (UBICS), Universitat de Barcelona, 08028, Barcelona, Spain}
\affiliation{ICREA, Pg. Llu\'{\i}s Companys 23, 08010, Barcelona, Spain}

\begin{abstract}
We introduce a model for the randomization of complex networks with geometric structure. The geometric randomization (GR) model assumes a homogeneous distribution of the nodes in an underlying similarity space and uses rewirings of the links to find configurations that maximize a connection probability akin to that of the $\mathcal{S}^1$ or $\mathcal{H}^2$ geometric network models. However, GR preserves the original degree sequence, as in the configuration model, thus eliminating the fluctuations of the degree cutoff. Moreover, the model does not require the explicit estimation of hidden degree variables, which restricts the number of free parameters to one, controlling the level of clustering in the rewired network. We illustrate the potential of GR as a null model by investigating the effects on modularity that derive from the flattening of geometric communities in both real and synthetic networks. As a result, we find that for real networks the geometric and topological communities are consistent, while for the randomized counterparts, the topological communities detected are attributable to structural constraints induced by the underlying geometric architecture.
\end{abstract}

%\date{\today}
\maketitle

\section{Introduction} 
Null models play a central role in network science and statistics to discern regularities and patterns in the fabric of systems that are not attributable to specific constrains. Typically, null models of complex networks are fit with one or several particular structural properties, depending on the question at hand, to predict the organization of a network as the outcome of a random process where other features are allowed to vary. Hence, null models are said to produce maximally random ensembles given some specific features~\cite{Newman:2001zb}. Many successful applications of null models in complex networks include the detection of rich-club ordering~\cite{Colizza:2006zw,Serrano:2008zw}, the characterization of structural correlations in weighted networks~ \cite{Garlaschelli:2009or}, or the quantification of communities using modularity~\cite{Newman:2004aa}.

Intriguingly, the frontier separating models and null models is not so neat, specially when the models remain simple and the null models fix more than one property. In fact, some famed network models, originally born to explain some peculiarity of the structure of networks on the basis of first principles, are often used as null models, for instance, the growing Barab\'{a}si-Albert model~\cite{Barabasi:1999ay} that explains the generation of scale-free degree distributions implementing a preferential attachment mechanism. Recently, a class of network models in hidden metric spaces~\cite{Serrano:2008ga,Boguna:2009uz} has been shown to explain many pivotal features of real networks simultaneously ---like the small world property, heterogeneous degree distributions, high levels of clustering, and self-similarity--- based only on three parameters, {controlling the average degree, the exponent of the power-law degree distribution and the clustering coefficient.}

The key ingredient of the geometric network models is the fact that the probability to connect two nodes of the network is determined by their effective distance, as measured in a hidden metric space in which nodes are embedded. 
The underlying space is defined along two dimensions representing popularity and similarity features of the nodes, such that more popular and similar nodes have more chance to interact.   
In the $\mathcal{S}^1$ model~\cite{Serrano:2008ga}, the hidden degree of a node is a proxy for its popularity, and its angular position in the one-dimensional sphere (or circle) provides the similarity measure. 
The two coordinates contribute explicitly to the connection probability between two nodes, which increases with the product of their hidden degrees and decreases with their angular distance along the circle.
The hidden degree can be estimated by the observed degree and reinterpreted as a radial coordinate in a {hyperbolic plane~\cite{KrPa10}, which leads to the formulation of an isomorphic version of the model which is purely geometric. In the $\mathcal{H}^2$ model, popularity takes the form of a radial coordinate in the hyperbolic disk, such that higher degree nodes are placed closer to the center, while the angular coordinate remains as in the $\mathcal{S}^1$ similarity space, and the probability of connection decreases with the hyperbolic distance. 

In both  $\mathcal{S}^1$ and $\mathcal{H}^2$ models, the angular coordinate of nodes, representing the similarity dimension, is extracted from a homogeneous distribution,  at odds with hyperbolic maps of real networks~\cite{Boguna2010}}. In fact, geometric communities of nodes lying nearby in {the similarity space (referred as soft communities or latent communities) are typically detected in real networks~\cite{Boguna2010,Serrano:2012we,garcia-perez:2016} and can be modeled~\cite{Zuev:2015aa,garcia-perez:2018aa}}.
This observation opens the door to the use of {geometric models with homogeneous similarity distribution} as null models for the investigation of the community organization and other structural properties of geometric networks. \\
 
In this paper, we introduce a variant of {the popularity-similarity} geometric model, that we named geometric randomization (GR) model, and illustrate its use as a null model for the analysis of the topological properties of real networks, including community structure.
The GR model assumes the same form of the connection probability as in the $\mathcal{S}^1$ or $\mathcal{H}^2$ models, and a homogeneous distribution for the similarity coordinate as well. 
In contrast, it is fit with a given degree-sequence, like the configuration model~\cite{newmanbook}. 
The use of prescribed degrees allows to skip the step of estimating the hidden degrees from real data. 
It could also help, for instance, in the analysis of features which are specially sensitive to fluctuations of the degree cutoff, like the behavior of  dynamical processes such as epidemic spreading or synchronization, or for high-fidelity reproduction of real network topologies. 
Based on the premises mentioned above, we propose an algorithm that homogenizes the similarity distribution and rewires the links in a network preserving the given degrees to maximize the likelihood that the new topology is generated by the geometric model. 
We analyze the effects of the GR model on the topological properties of real and synthetic geometric networks, and use it as a null model to explore the effects on modularity of the flattening of geometric communities in the similarity space. 

\section{The geometric randomization model}

The GR model operates on networks where nodes have an observed degree and exist in a similarity space. The similarity space is taken to be a circle, as in the $\mathcal{S}^1$ or $\mathcal{H}^2$ models. In those models every node $i$ is characterized by a popularity-similarity pair $(\kappa_i,\theta_i)$, where $\kappa_i$ is the node's hidden degree (expected to be proportional to the observed degree $k_i$) and $\theta_i$ its angular or similarity coordinate.\\
In the GR model, instead, only angular coordinates are assigned to the nodes, chosen uniformly at random from $[0,2\pi]$. The network is then rewired in order to maximize the likelihood that the new topology is generated by the $\mathcal{S}^1$ model while preserving the observed degrees, and thus the total number of edges $E$. 
The rewiring procedure is conducted by executing the Metropolis-Hastings algorithm,
 aimed at finding the network connectivity (i.e. the adjacency matrix $a_{ij}$) that maximizes the likelihood function
\begin{equation}
\label{eq:likelihood}
\mathcal{L} = \prod_{i<j} p(\kappa_i,\kappa_j,\Delta\theta_{ij})^{a_{ij}} \left[(1-p(\kappa_i,\kappa_j,\Delta\theta_{ij}))^{1-a_{ij}} \right],
\end{equation}
where $\Delta\theta_{ij}$ stands for the angular distance between nodes $i$ and $j$, and the $\mathcal{S}^1$ connection probability $p(\kappa_i,\kappa_j,\Delta\theta_{ij})$ reads
\begin{equation} \label{eq:conn_prob}
  p(\kappa_i,\kappa_j,\Delta\theta_{ij}) = \frac{1}{1 + \left( \frac{\Delta\theta_{ij}R}{\mu \kappa_i \kappa_j} \right)^\beta} = \frac{1}{1+ \chi_{ij}^{\beta}}=\tilde{p}(\chi_{ij}).
\end{equation}
Parameter $\mu$ depends on the observed average degree $\langle k \rangle$ of the network, and $R$ is the radius of the circle (adjusted to have a density of nodes equal to 1, see Appendix A) .

The algorithm proceeds by repeating the following steps:
\begin{itemize}
\item Compute the current likelihood $\mathcal{L}_c$
\item Two links, between nodes $i$ and $j$, and between nodes $l$ and $m$, are randomly chosen and swapped: the new links are connecting nodes $i$ and $m$, and nodes $l$ and $j$.
\item Compute the new likelihood $\mathcal{L}_n$
\item If $\mathcal{L}_n > \mathcal{L}_c$ accept the link swap
\item Otherwise, if $\mathcal{L}_n < \mathcal{L}_c$ accept the link swap with probability $\mathcal{L}_n / \mathcal{L}_c$ 
\end{itemize}
The rewiring algorithm is terminated after a number $E^2$ of edges are chosen to be swapped, ensuring that the likelihood has reached a plateau. Notice that at the end of the rewiring procedure the degrees of the nodes have not changed but the resulting network might not be connected. Since the hidden degrees are kept constant (independently of their values), the probability of swapping links 
between nodes $i$ and $j$ and between nodes $l$ and $m$ simply reads 
\begin{equation}
\label{eq:likelihood_diff}
\mathcal{L}_n / \mathcal{L}_c = \left( \frac{ \Delta \theta_{ij} \, \Delta \theta_{lm} }{ \Delta \theta_{il} \, \Delta \theta_{jm}}  \right)^\beta.
\end{equation}
Therefore, the GR model does not actually require to estimate the hidden degrees of the nodes because they do not enter in any step of the algorithm. In contrast, the GR model simply needs to assign uniformly distributed angular coordinates and give a value for the clustering parameter $\beta$, see next Section for details on this part. \\

Geometric randomizations of networks can be also obtained using the $\mathcal{S}^1$ model with parameters $\gamma$, $\beta$ and $\mu${---controlling the exponent of the power-law hidden degree distribution, the clustering coefficient, and the average degree, respectively---} estimated from the empirical network. This alternative however, requires the explicit estimation of the hidden degree sequence $P(\kappa)$ or of the exponent of the hidden degree distribution, and, thus, it may introduce undesired fluctuations in the degree cutoff which can induce relevant differences between the topological properties of real and $\mathcal{S}^1$ generated networks.

\section{Tuning clustering through parameter $\beta$}

In order to apply the GR model to a real or synthetic network one simply needs to fix parameter $\beta$, which controls the level of clustering in the network~\cite{Serrano:2008ga}.  Clustering is a signature of the metricity of geometric networks~\cite{Krioukov:2016} and gives the connection between the observed topology and the underlying metric space, as a reflection of the triangle inequality.

Note that the value of $\beta$ affects the probability to accept a link swap (see Eq.~\eqref{eq:likelihood_diff}) so it determines the final network's structure. We address the role of $\beta$  by applying the GR model to synthetic networks generated by the Geometric Preferential Attachment (GPA) model~\cite{Zuev:2015aa} and the soft communities in similarity space (SCSS) model~\cite{garcia-perez:2018aa}. Both models are intended to produce synthetic networks with tunable community structure.

 The GPA model generates geometric networks with soft-communities using a growing mechanism in the hyperbolic plane.  
%in which nodes appear at a certain time with a certain radial and an angular coordinate, $r_i(t)$ and $\theta_i(t)$, and connect to the $m$ hyperbolically closest neighbors.
The probability of connection depends on parameter $\Lambda$ controlling the initial attractiveness of the different angular regions, such that the heterogeneity of the angular coordinate is a decreasing function of $\Lambda$, with $\Lambda \rightarrow \infty$ recovering the homogeneous distribution. 
Notice that the degree distribution and the clustering coefficient in networks generated by the GPA model are independent of $\Lambda$. 
However, $\beta \rightarrow \infty$ by construction and, thus, the level of clustering is always the maximum possible. 
The SCSS model consists in an $\mathcal{S}^1$ version for the generation of soft communities that allows to change the generated level of clustering as a function of $\beta$.\\

\begin{figure}[tbp]
  \begin{center}
    \includegraphics[width=0.5\textwidth]{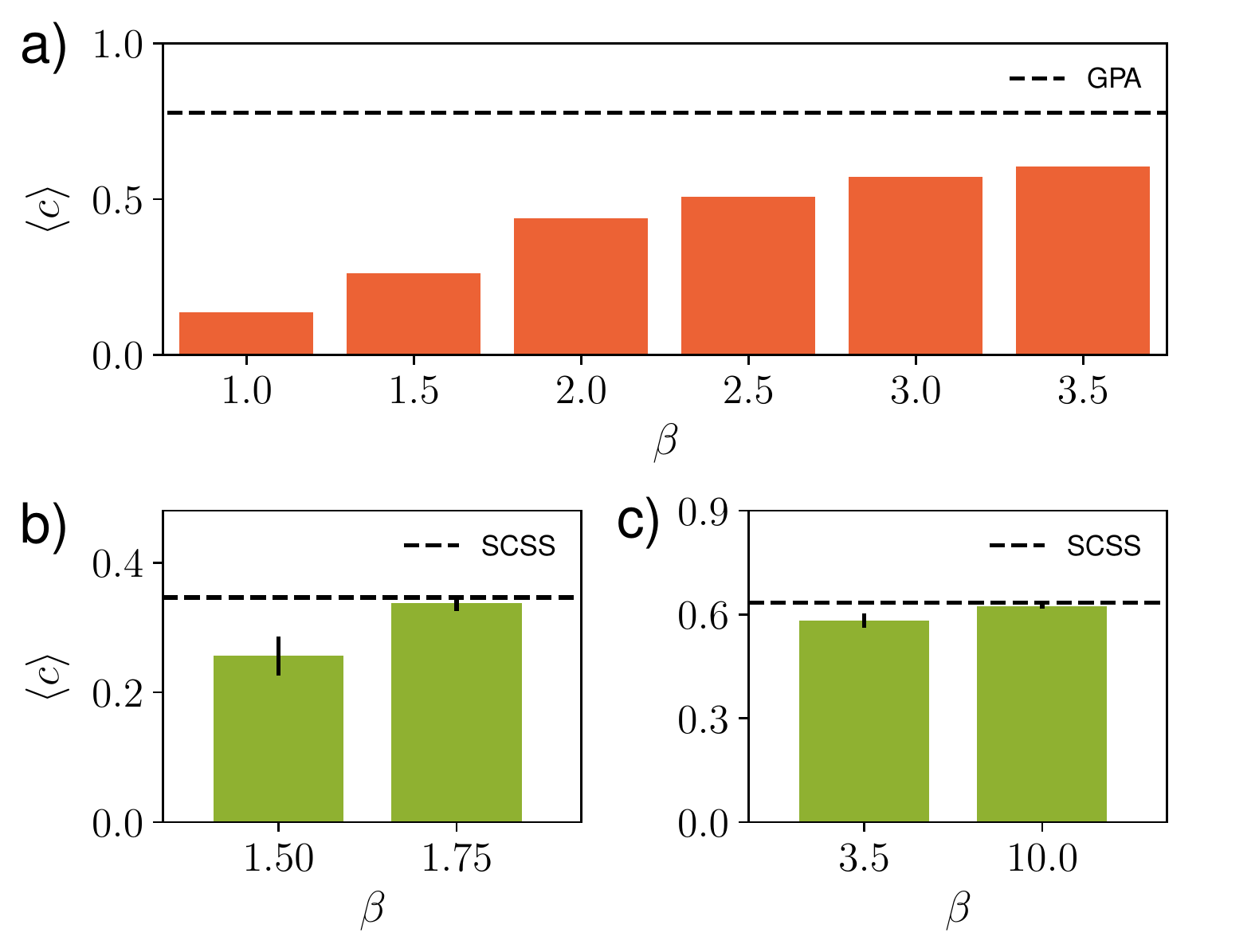}
  \end{center}
  \caption{\textbf{a) Average clustering $\av{c}$ of a network generated by the GPA model} (dashed line) and rewired versions (orange) obtained by applying the GR model with different values of $\beta$. The networks have size $N = 10^3$, exponent of the degree distribution $\gamma=2.5$, number of links per node $m=4$, and the initial attractiveness $\Lambda=0.1$. \textbf{b) Average clustering of two networks generated with the SCSS model} (dashed line) with attractiveness $\Lambda=0.1$ and $\beta_0=1.5$ in b) and $\beta_0=3.5$ in c). Green bars indicate the $\langle c \rangle$ of networks obtained by applying the GR with $\beta_0$ and with $\beta$, respectively.}
  \label{Fig1}
\end{figure}

Fig.~\ref{Fig1}a shows the average clustering coefficient $\av{c}$ of a GPA network compared with the randomizations obtained by applying the GR model using different values of $\beta$. 
As expected, the average clustering of the rewired networks strongly depends on the value of $\beta$: the lower $\beta$, the lower $\av{c}$ in the resulting network. A level of clustering similar to GPA values can be obtained in GR networks by using large values of $\beta$, such as $\beta=10$.

In Fig.~\ref{Fig1}b-c, we report the average clustering coefficient obtained by applying the GR model to synthetic networks generated with the SCSS model. The SCSS networks are produced using two different generating values, referred as $\beta_{0}$.
Fig.~\ref{Fig1}b-c show that it is possible to fine tune the value of $\beta$ used by the GR networks so that they reproduce the same average clustering $\av{c}$ as the original networks.
If the generation value $\beta_0$ is used for the rewiring, the level of clustering in the GR instances does not reach that in the original networks and remains smaller. 
This observation can be understood by noticing the following two points. First, for SCSS networks the $\av{c}$ is independent of the level of angular clusterization, so any two SCSS networks with equal $\beta_0$ and the same distribution of hidden degrees, $P(\kappa)$, will have equal $\av{c}$. Second, a GR instance of a SCSS network obtained using $\beta_0$ would be one with homogeneous $P(\theta)$ and the same observed degree distribution $P(k)$ as in the SCSS network. That is, if $P(k)=P(\kappa)$ exactly, then the average clustering $\av{c}$ reached by the GR instance with $\beta_0$ would need to match that of the SCSS network. Since we do not observe this matching in Fig.~\ref{Fig1}b-c, we conclude it is due to differences between the distribution of observed and hidden degrees of the SCSS network.
%\elis{MAYBE : Add some last sentece llinking this to the importance of not having to estimate hidden degrees in real nets as in our null model; coz as we've seen if they are not very well estimated this can induce difficulties in reproducing/keeping invariable some topological properties.  }\\

\section{Effects of geometric randomization in empirical networks}

\setlength{\tabcolsep}{0.40em}
{\renewcommand{\arraystretch}{1.25}% for the vertical padding 
\begin{table}[tbp]
  %\begin{ruledtabular} 
    \begin{tabular}{|l| c c c c c c c|}
    \botrule
      \textbf{Data set}     & $N$     & $\gamma$  & $\beta_0$ &  $\beta$  & $\av{k}$ & $\av{c}$ & $ D_{KS} $ \\ \hline
      Enron        &  33696  & 2.14      &  2.70   &  2.60       & 10.73    & 0.71     & 0.027\\ \hline % 1.47292753076e-09
      Comms.  &  374    & 2.50      & 1.06    & 1.25        & 5.83     & 0.22     &   0.144 \\ \hline%0.00256772101877 
      Metabolic    &  1436   & 2.60      &  2.13   & 2.50        &  6.57    & 0.54     &   0.092 \\ \hline % 5.47021703629e-06 
      Words        &  7377   & 2.25      & 1.01    & 1.00        & 11.98    & 0.47     &   0.116 \\\hline  %1.57183247515e-42 
      Internet     &  23748  & 2.16      & 1.88    &  2.20       & 4.92     & 0.61     & 0.123 \\ \hline    %5.81170331553e-143 
      Music        &  2476   & 2.27      &  2.50   & 2.65        & 16.66    & 0.82     &  0.072\\ \botrule %1.05251207708e-07 
    \end{tabular} 
  %\end{ruledtabular} \pm
    \caption{\textbf{Properties of the data sets under consideration}: $N$, size of the network; $\gamma$, exponent of the power-law form fitting the degree distribution, $P(k) \sim k^{-\gamma}$; parameter $\beta_0$ estimated from the embedding of the real network; parameter $\beta$ that preserves the level of clustering in the GR network; $\av{k}$, average degree; and the $D$  score ($95\%$ CI) of the KS test performed between the $P(\theta)$ distributions of the original networks and networks obtained by applying the GR model (see main text). } \label{tab:summary} 
\end{table}
}

In the following, we apply the GR model to real networks. 
We consider six empirical networks from different domains:  the network of chords transitions {in western popular music (Music)~\cite{Serra:2012}, the one-mode projection onto metabolites of the human metabolic network at the cell level (Metabolic)~\cite{Serrano:2012we}, 
the word adjacency network in Darwin's book {\it On the Origin of Species}} (Words)~\cite{Serrano2009Words}, the email communication network within the Enron company (Enron)~\cite{klimt:2004},
and the Internet at the autonomous system level (Internet)~\cite{Claffy:2009fe,Boguna2010}, see Table \ref{tab:summary} and Appendix B for details. 

\begin{figure}[tbp]
  \begin{center}
    \includegraphics[width=0.48\textwidth]{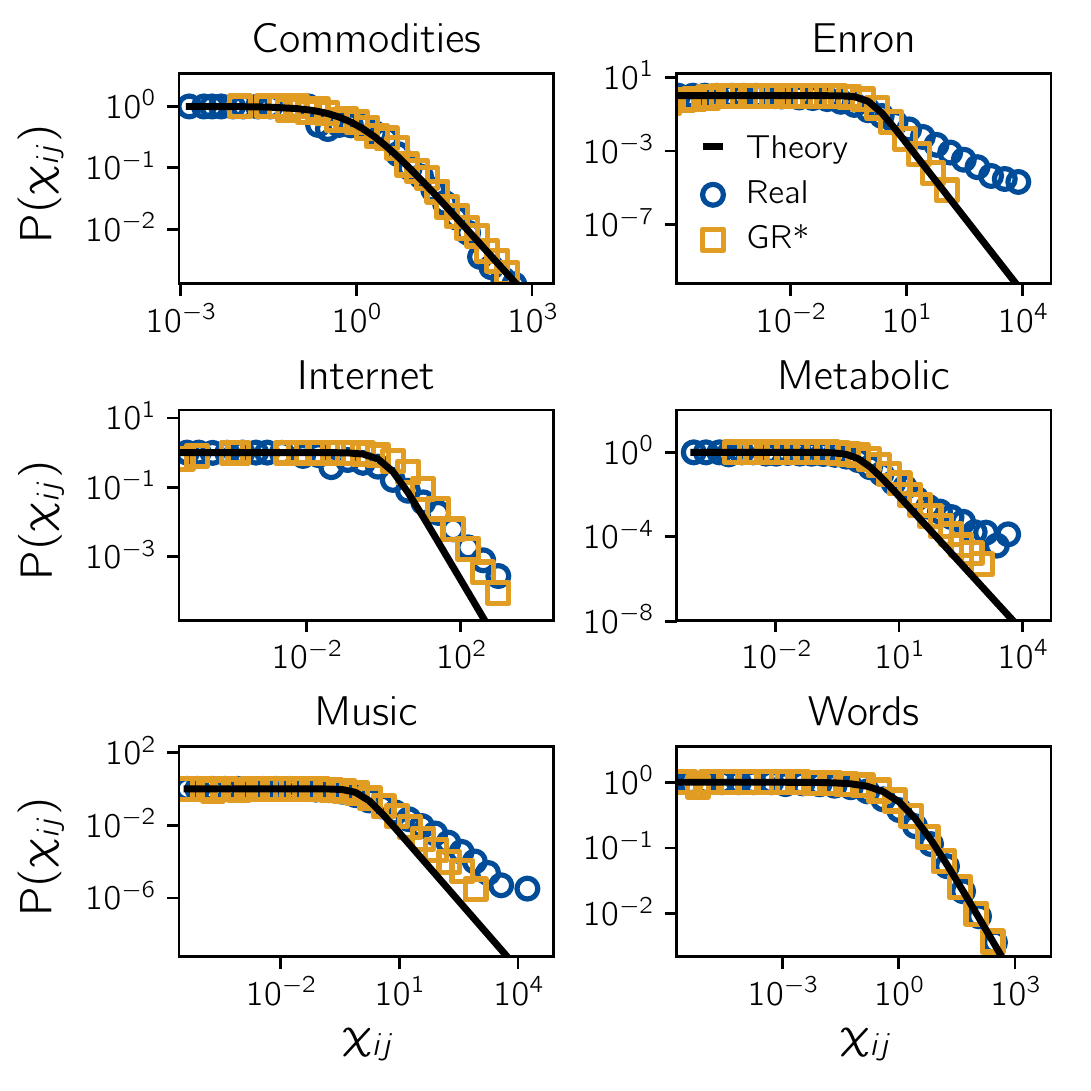}%GPA_l01_clus.pdf
  \end{center}
  \caption{\textbf{Empirical connection probability} for original (blue dots) and GR (orange dots) networks. Fraction of connected pairs of nodes as a function of $\chi_{ij}=\Delta\theta_{ij} R/(\mu \kappa_i \kappa_j)$. The black line shows the theoretical curve,  Eq.~(\ref{eq:conn_prob}).} 
  \label{Fig2}
\end{figure}

As described in the previous Section, $\beta$ is the only free parameter of the model, and can be used to tune the clustering coefficient. In the following, we will show results by using  a value of $\beta$ ensuring that the average clustering of the rewired network is equal to that of the real one. Another possible choice for $\beta$  is the value estimated when embedding the real network into the underlying metric space~\cite{Boguna2010}, which we indicate as $\beta_0$ in Table \ref{tab:summary}.
The embedding method estimates the coordinates of the nodes in the underlying geometry by maximizing the likelihood that the observed topology has been produced by the model.  In the process, $\beta_{0}$ is estimated such that the expected clustering coefficient of the embedded network matches the observed clustering coefficient of the network topology. As explained in the previous section for synthetic networks, using $\beta_0$ as the input in GR does not produce in general rewired networks with the same average clustering $\av c$ as in the original networks.  
For real networks, the two values of $\beta$ are very similar but not always identical, see Table \ref{tab:summary}.
The small difference is related with the fact that, for some real networks, the GR model cannot adjust simultaneously the empirical connection probability and the observed clustering using a single value of $\beta$, see Fig.~\ref{Fig2}. \\

\subsection{Clustering and degree correlations}

\begin{figure}[tbp]
  \begin{center}
       \includegraphics[width=1.00\columnwidth]{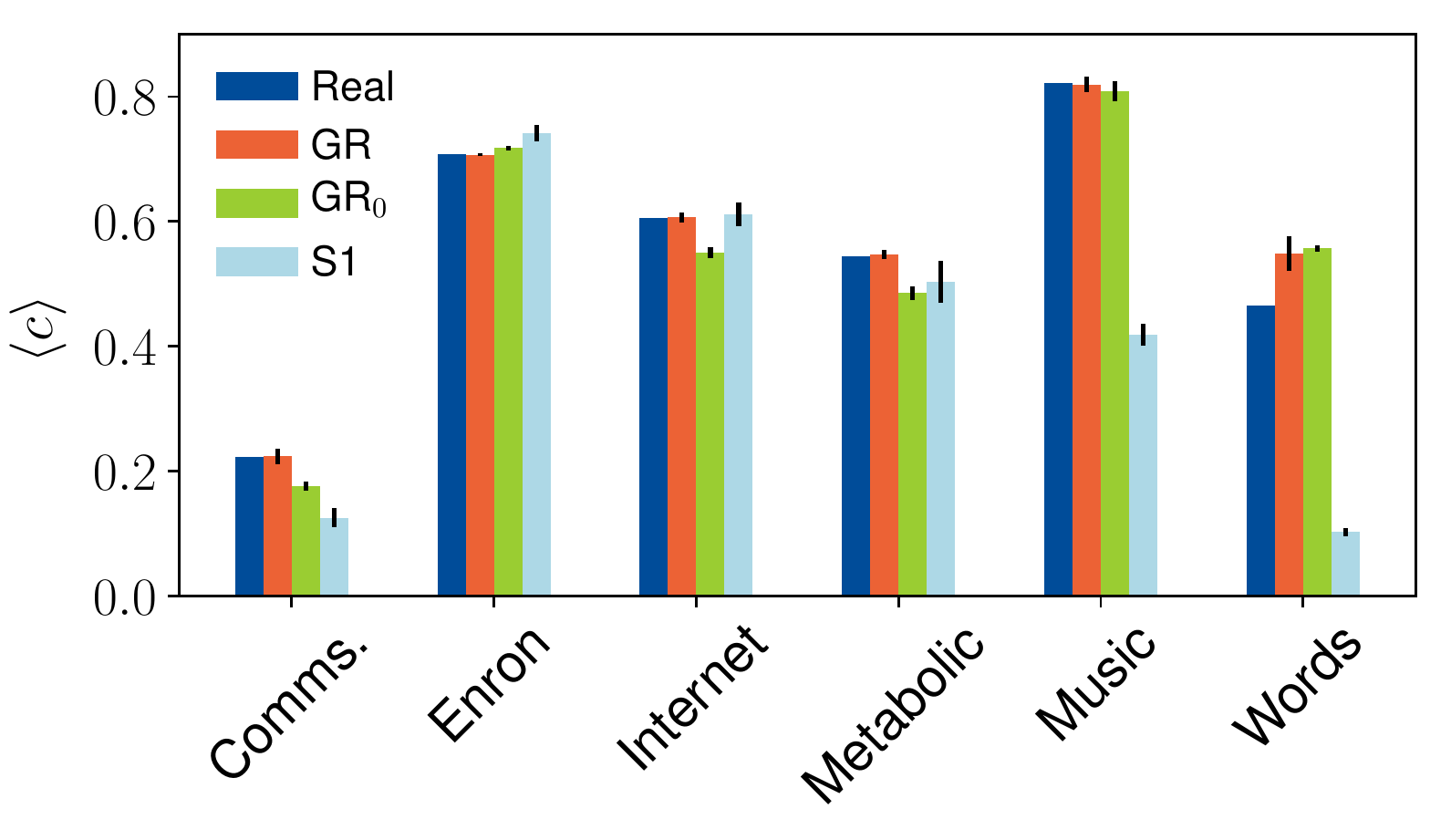}
  \end{center}
  \caption{\textbf{Average clustering} $\av{c}$ of  empirical networks (blue), networks obtained from the GR (red) and S1 (light blue) models.   
  GR networks obtained with $\beta_0$ (green) are indicated as GR$_0$. Error bars are calculated over 10 realizations of the GR and S1 models.  }    \label{Fig3}
\end{figure}

\begin{figure*}[tbp]
    \includegraphics[width=1.01\textwidth]{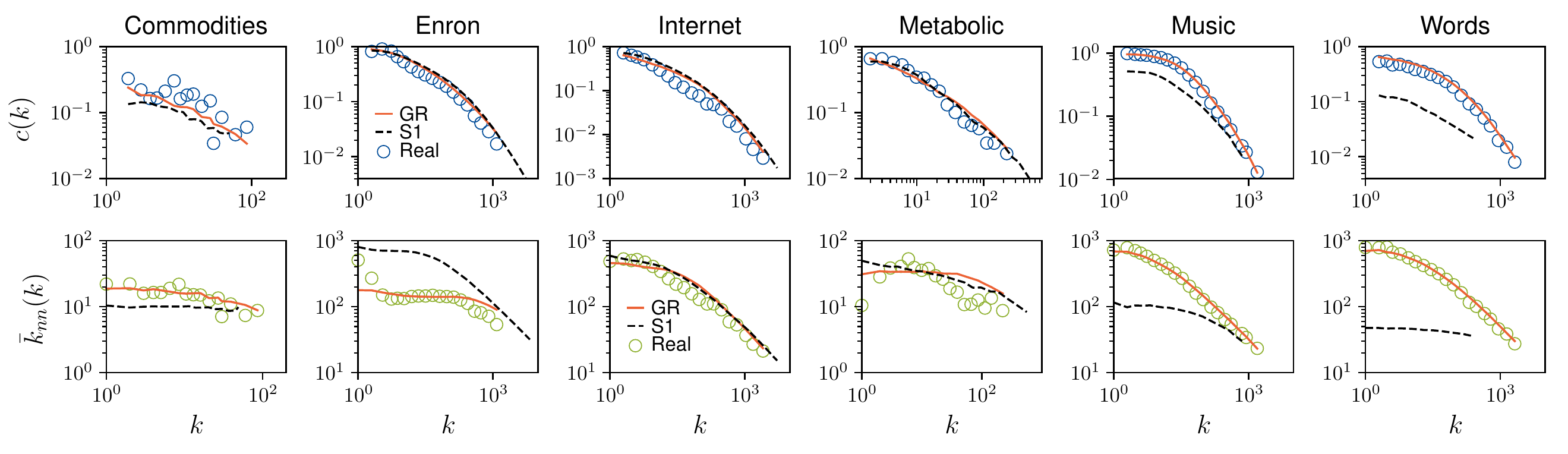}
    \caption{ \textbf{Clustering $c(k)$ (top) and average degree of nearest neighbors $\bar{k}_{nn}(k)$  (bottom) } as a function of the degree,  for empirical networks (dots), and networks obtained from the GR (continuous orange line) and S1 (black dashed line) models. }
  \label{Fig4}
\end{figure*}

Fig.~\ref{Fig3} shows the average clustering $\av{c}$ of the empirical  networks under consideration as compared to the randomized versions obtained by the GR model. We consider both values $\beta$ and $\beta_0$ (the corresponding networks are indicated by GR and GR$_0$, respectively), and we include also a comparison with real network replicas generated by the S1 model~\cite{Serrano:2008ga}, see Appendix A. As expected, GR networks show an average clustering practically identical to that of the original data, while GR$_0$ networks present mild deviations, and differences are usually more important for S1 networks due to deviations in the obtained degrees.  
One exception to the preservation of clustering in GR instances is the Words data set. This empirical network has a $\beta_0$ extremely close to the minimal threshold of $\beta_0=1$ defined in hidden metric space network models. The $\beta$ value necessary to ensure that the GR network has the same level of clustering as the empirical one cannot be achieved since it would need to be lower than 1. In general, an embedding value of $\beta_0\simeq1$ suggests that clustering is due to finite size effects, since $\beta_0=1$ corresponds to absence of clustering in the thermodynamic limit of the geometric network models.

Graphs on the top row of Fig.~\ref{Fig4} show the clustering spectrum $c(k)$ for empirical networks and networks obtained by the GR  and S1 models. In all cases, the functional form of $c(k)$ is similar, a decreasing function of $k$ with a broad tail. The clustering spectrum of the GR networks is always very close to the original data, while the S1 networks present important departures in some systems, as a result of the lack of preservation of the empirical degrees. This is especially evident for the S1 versions of the Music and Words networks, with the clustering spectrum much lower than that of the original data.  

On the other hand, the real networks under consideration are generally disassortative, as revealed by the decreasing form of the  
average degree of nearest neighbors, $\bar{k}_{nn}(k)$ function, Fig.~\ref{Fig4} (bottom). Internet, Music and Words show a decay with power law form, while other data sets show milder degree correlations. In all cases, GR networks have $\bar{k}_{nn}(k)$  distributions very similar to the original data, while S1 networks exhibit strong deviations, with the exception of the Internet.
\newline

\subsection{Community structure}

\begin{figure*}[tbp]
  \begin{center}
    \includegraphics[width=2.00\columnwidth]{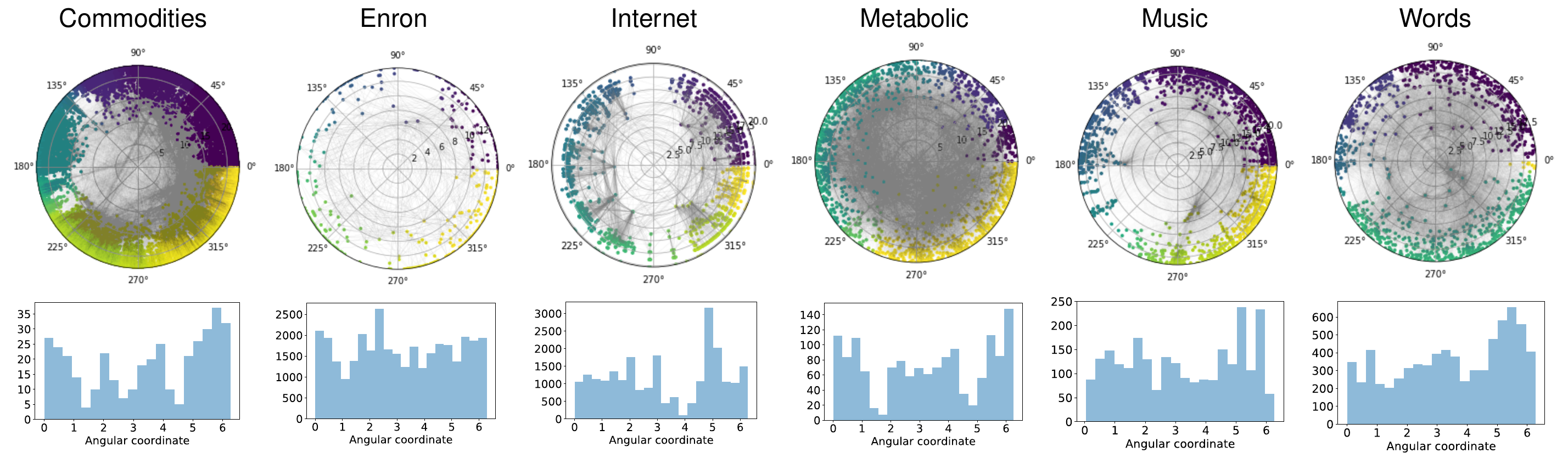}
  \end{center}  
  \caption{ \textbf{Top row}: Empirical networks embedded in the hyperbolic disk. Distinct communities are indicated by different colors. 
  \textbf{Bottom row}: Probability distribution of the angular coordinate, $P(\theta)$, of the empirical networks. }
  %Data sets plotted are, from left to right: Enron, Metabolic, Commodities, Words, Internet, Music. }
  \label{Fig5}
\end{figure*}

So far, GR randomized versions of real and synthetic geometric networks seem to be able to preserve topological features beyond the degree distribution, including clustering and the average nearest neighbors degree.
However, the GR randomization homogenizes the distribution of nodes in similarity space, while nodes in real networks are typically heterogeneously distributed,  as they are more concentrated in some specific regions~\cite{Serrano:2012we,garcia-perez:2016}. 
This denotes the presence of communities of similar nodes, named soft communities~\cite{Zuev:2015aa}. 
Top row of Fig.~\ref{Fig5} shows the representations of the empirical networks embedded in the hyperbolic plane, with coordinates $(r, \theta)$ (see Appendix A for the relationship between $r$ and the degree, and Appendix B for references to the sources of the empirical maps). 
One can clearly see that the angular coordinates $\theta$ are heterogeneously distributed in $[0, 2\pi]$. 
A different perspective is shown in the bottom row in Fig.~\ref{Fig5}, displaying the probability density function $P(\theta)$ of the similarity coordinate of the nodes for the six empirical networks.  

The heterogeneity of the angular coordinate can be quantified by performing a Kolmogorov-Smirnov (KS) test between the probability density functions $P(\theta)$ and $P_{GR}(\theta)$. 
The KS statistic measures the difference between two probability distributions, and it is defined as the maximum difference
between the values of the distributions $P(\theta)$ and $P_{GR}(\theta)$. The larger the KS score, the more heterogeneous the angular distribution. Thus, it can be used to discard the null hypothesis that the empirical $P(\theta)$ and synthetic $P_{GR}(\theta)$ samples (with uniform distribution by construction) present the same angular distribution. The KS distance $D_{KS}$ for empirical networks under consideration is reported in Table \ref{tab:summary}. 
One can see that the null hypothesis is strongly rejected for all real networks. 

Soft communities in the geometric domain can then be detected using geometric methods. We use the definition of soft communities given in~\cite{Zuev:2015aa}, where they are defined as group of nodes in similarity space separated from the rest by two angular gaps that exceed a certain critical value, $\Delta \theta_c$. The critical gap $\Delta \theta_c$ is calculated as the expected value of the largest gap between two nodes when the angular coordinates are distributed uniformly at random: $\Delta \theta_c \simeq  2 \pi \ln(N)/N$. In the top row of Fig.~\ref{Fig5}, we highlight the soft community deterministic partition detected by the critical gap method in the real networks using different colors.

\begin{figure}[tbp]
  \begin{center}
    \includegraphics[width=1.00\columnwidth]{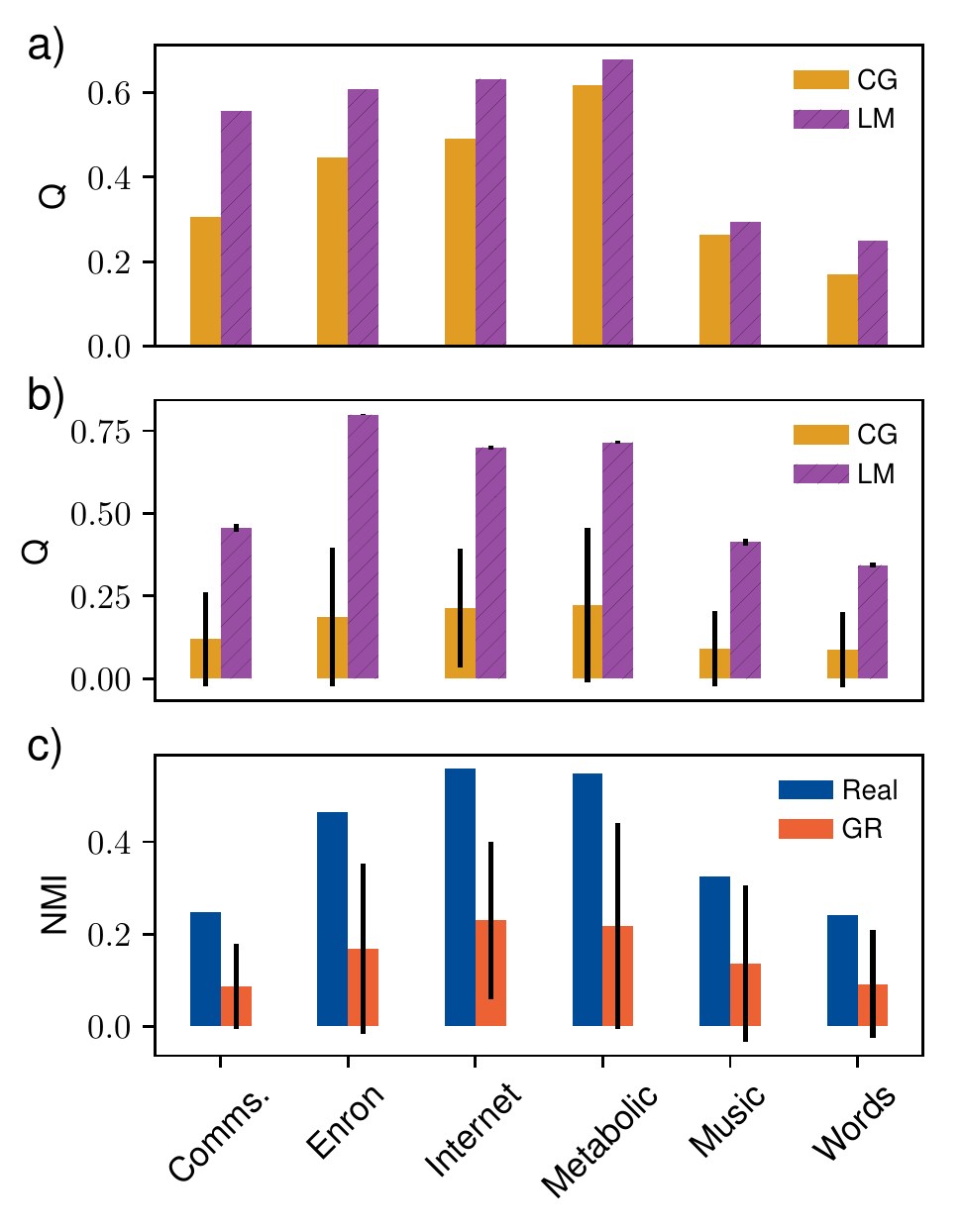}
  \end{center}  
  \caption{ \textbf{a-b) Modularity $Q$} as detected by the Lovain method (purple) and the critical gap (yellow), for real (plot a)) and GR (plot b)) networks.  Error bars in plot b) are obtained by 10 realizations of the GR model. \textbf{c)  Normalized mutual information} between the partition detected by the Louvain and the critical gap methods, for empirical (blue) and GR (red) networks.  Error bars are obtained by 10 realizations of the GR model.  }
  \label{Fig6}
\end{figure}

Next, we compare the community structure of the real networks with their randomized counterparts. To quantify their topological community structure, we apply the widely used Louvain method~\cite{Blondel:2008}, aimed at maximizing the modularity $Q \in [-1,1]$, that compares the fraction of links inside communities with the expected fraction for a random distribution of edges with the same node degree distribution as the given network.
%Figs.~\ref{Fig6}a show a comparison of the modularity of the community structure detected by the critical gap \eo{(CG)}, with the one identified by the Louvain method \eo{(LM)}, for both original and GR networks. 
Interestingly, Fig.~\ref{Fig6}a shows that in real networks, albeit the Louvain method identifies topological communities with higher modularity, the soft communities discovered by the CG display large $Q$ values, in some cases (e.g. Metabolic or Music data sets) comparable to the modularities given by the purely topological LM.

This picture is completely different for GR networks,  reported in Fig.~\ref{Fig6}b. 
GR networks show strong community organization at the topological level, resulting in large values of $Q$ as measured by the Louvain method, which is induced by structural constraints imposed by the geometric models~\cite{Radicchi2018}. 
However, as expected, the critical gap does not detect soft communities, as demonstrated by the non-significant values of the modularity, compatible with zero, over different realizations of the randomization process.  

We study in more detail the relationship between soft communities and topological ones by comparing the partition obtained by the Louvain method with the partition generated by the critical gap. 
The overlap between the two partitions can be quantified by the normalized mutual information~\cite{Cover:1991jl}. Fig.~\ref{Fig6}c shows that the overlap between geometric and topological communities is quite large for real networks, specially for Metabolic and Internet data sets, meaning that communities identified by purely (deterministic) geometric methods are meaningful, though subject to the degree of congruency of the real network with the hidden metric space. On the contrary, Fig.\ref{Fig6}c shows that the overlap between soft and topological communities in GR networks is very low due to the complete randomization of the angular coordinate operated by GR. 

\section{Conclusions}

The rewiring process preserving degrees in the geometric randomization of real networks gives an alternative to their replication using directly the popularity-similarity model as a topology generator. The GR offers the advantage of avoiding the delicate task of estimating the hidden degree distribution, and it can be especially useful in problems responsive to fluctuations of the degree cutoff, like the behavior of some dynamical processes including epidemic spreading processes.

As a model, GR depends on a single parameter controlling the level of clustering in the resulting networks, so that the clustering coefficient of real networks can be chosen to be replicated or not. Interestingly, the discrepancies between hidden and observed degrees in embedded networks, have an effect on the clustering level achieved by the GR. In particular, the parameter value suggested by the embedding of the original data is, in general, not far but not totally coincident with the needed  value for replicating the clustering coefficient of the original network. Our results also indicate that, in some networks, degree-degree correlations can only be replicated by the geometric network models if the observed degrees are preserved.

As a null model, GR can be used to investigate the relevance of geometric communities in real networks. Taken together, our results indicate that geometric communities are meaningful in the real networks analyzed here.  At the same time, topological communities, like those detected in GR networks, are not always reliable and can be a result of constraints induced by the underlying geometric architecture. The fact that an underlying geometric organization imposes structural constraints on complex networks, which are strong enough for recreating  detectable topological communities even in the absence of geometric ones, is an interesting subject by itself and will be investigated in future work.

\section{Acknowledgements}
We thank Mari\'an Bogu\~{n}\'a and Guillermo Garc\'{\i}a-P\'erez for helpful discussions. We acknowledge support from a James S. McDonnell Foundation Scholar Award in Complex Systems; Ministerio de Ciencia, Innovaci\'on y Universidades of Spain project no. FIS2016-76830-C2-2-P (AEI/FEDER, UE); and the project {\it Mapping Big Data Systems: embedding large complex networks in low-dimensional hidden metric spaces} -- Ayudas Fundaci\'on BBVA a Equipos de Investigaci\'on Cient\'{\i}fica 2017.

\appendix*
\section{Appendix A. The $\mathcal{S}^1$ and $\mathcal{H}^2$ models} 
In the $\mathcal{S}^1$ model~\cite{Serrano:2008ga}, every node is characterized by hidden degrees and angular coordinates $(\kappa_i,\theta_i)$ representing the popularity (related to the degrees), and similarity dimensions. The $N$ nodes of the network are distributed at random in the similarity space, which is taken to be a one-dimensional sphere or circle of radius $R_{\mathcal{S}^1}=N/2\pi,$ adjusted to have a density of nodes equal to 1. Every pair of nodes is connected with a probability 
\begin{align} \label{eq:probability_of_connection_S1}
  p_{ij} = \frac{1}{1 + \left( \frac{\Delta\theta_{ij}R_{\mathcal{S}^1}}{\mu \kappa_i \kappa_j} \right)^\beta},
\end{align}
where $\Delta\theta_{ij}$ stands for the angular separation between nodes $i$ and $j$ in the similarity circle, and the parameters $\mu$ and $\beta$ control the average degree of the network and the level of clustering, respectively.

There exists an isomorphism between the $\mathcal{S}^1$ model and a version in hyperbolic space, the $\mathcal{H}^2$ model~\cite{KrPa10}, where the hidden degrees $\kappa$ are transformed into a radial coordinate, $r$, in a hyperbolic disk of radius $R_{\mathcal{H}^2}$ such that
\begin{align}
  \kappa \sim e^{(R_{\mathcal{H}^2} \ - \ r)/2} \ .
\end{align}
Consequently, nodes closer to the center of the hyperbolic disk have a higher expected degree and every node $i$ has then a radial and an angular coordinate $(r_i,\theta_i)$. A link between two nodes $i$ and $j$ exists with a probability $p(d_{ij})$ that depends on their distance $d_{ij}$, measured in the hyperbolic hidden metric space, such that nodes with higher probabilities of being connected are closely positioned in that space. Therefore, the connection probability must be a decreasing function of distance between nodes and, specifically, it can be chosen to be
\begin{equation}
\label{eq:conn_prob_hyp}
p(d_{ij}) = \frac{1}{1 + \exp[\beta (d_{ij}-R_{\mathcal{H}^2})]},
\end{equation}
where the parameter $\beta$ still controls the network's clustering coefficient. The distance $d_{ij}$ in the hyperbolic plane is calculated using the hyperbolic law of cosines,
\begin{equation}
\mathrm{cosh}(d_{ij})=\mathrm{cosh}r_i\mathrm{cosh}r_j - \mathrm{sinh}r_i\mathrm{sinh}r_j\mathrm{cos}\Delta\theta_{ij},
\end{equation}
where $\Delta\theta_{ij}$ is the minimum angular distance between nodes $i$ and $j$.\\ 

To produce replicas of the real networks using the $\mathcal{S}^1$ model, we extracted the parameters from the empirical networks, namely the size $N$ and the exponent $\gamma$ of the degree distribution, and used the exponent $\beta_0$ given by the embedding of the network into the hyperbolic disk. In order to generate the hidden degree sequence $P(\kappa)$ we adjusted parameter $\mu$ to obtain the observed average degree $\av{k}$, see Table I.
\newline

\section{Appendix B. Empirical data sets.}

\noindent\textbf{US Commodities.} This network represents the flows of goods and services exchanged (in USD) among industrial sectors in USA during year 2007. The hyperbolic embedding was obtained from Ref.\cite{Allard:2017}.\\
\textbf{Enron.} It is the network of email messaging activity within employees from the Enron company. We use the network obtained in Refs.\cite{klimt:2004,leskovec:2009} and the hyperbolic embedding constructed in Ref.\cite{garcia-perez:2018}\\
\textbf{Internet.} This network consists of the connectivity data of the Internet at the autonomous systems level collected by the Archipelago project\cite{Claffy:2009fe} during June 2009 and embedded in hyperbolic space in Ref.\cite{Boguna2010}.\\
\textbf{Human metabolic.} This network is the one-mode projection of metabolites of the bipartite metabolic network of human cell metabolisms, as spatially embedded in Ref.\cite{Serrano:2012we}.\\
\textbf{Music.} In this network nodes are chords--sets of musical notes played in a single beat and links represent observed transitions among them, see Ref.\cite{Serra:2012}. We use the hyperbolic embedding of a sparser and undirected version of such network as reconstructed in Ref.\cite{garcia-perez:2018}.\\ 
\textbf{Words.} This is the network of adjacency between words in the book "The Origin of Species" by Darwin, see Ref.\cite{alon-super-networks}. We use the embedding presented in Ref.\cite{garcia-perez:2018}.

\bibliography{ref_full}

\end{document}